\documentclass[twocolumn,showpacs,preprintnumbers,amsmath,amssymb]{revtex4-1}
\usepackage{amsfonts}
\usepackage{amsmath}
\usepackage{amssymb}
\usepackage{graphicx}
\usepackage{dcolumn}
\usepackage{bm}
\usepackage{longtable}
\begin{document}
\title{Spatial effects in real networks: measures, null models, and applications}
\author{Franco Ruzzenenti}
\email[]{ruzzenenti@unisi.it}
\affiliation{Center for the Study of Complex Systems, University of di Siena, Via Roma 56, 53100 Siena (Italy)}
\affiliation{Department of Chemistry, University of di Siena, Via Aldo Moro 1, 53100 Siena (Italy)}

\author{Francesco Picciolo}
\affiliation{Department of Chemistry, University of di Siena, Via Aldo Moro 1, 53100 Siena (Italy)}

\author{Riccardo Basosi}
\affiliation{Department of Chemistry, University of di Siena, Via Aldo Moro 1, 53100 Siena (Italy)}
\affiliation{Center for the Study of Complex Systems, University of di Siena, Via Roma 56, 53100 Siena (Italy)}

\author{Diego Garlaschelli}
\affiliation{Instituut-Lorentz for Theoretical Physics, Leiden Institute of Physics, University of Leiden, Niels Bohrweg 2, 2333 CA Leiden (The Netherlands)}
\date{\today}
\begin{abstract}
Spatially embedded networks are shaped by a combination of purely topological (space-independent) and space-dependent formation rules.
While it is quite easy to artificially generate networks where the relative importance of these two factors can be varied arbitrarily, it is much more difficult to disentangle these two architectural effects in real networks.
Here we propose a solution to the problem by introducing global and local measures of spatial effects that, through a comparison with adequate null models, effectively filter out the spurious contribution of non-spatial constraints.
Our filtering allows us to consistently compare different embedded networks or different historical snapshots of the same network. 
As a challenging application we analyse the World Trade Web, whose topology is expected to depend on geographic distances but is also strongly determined by non-spatial constraints (degree sequence or GDP). 
Remarkably, we are able to detect weak but significant spatial effects both locally and globally in the network, showing that our method succeeds in retrieving spatial information even when non-spatial factors dominate.
We finally relate our results to the economic literature on gravity models and trade globalization.
\end{abstract}
\pacs{89.65.Gh; 89.70.Cf; 89.75.-k; 02.70.Rr}
\maketitle
\section{introduction}
The growth of interest towards complex networks during the last decade was mainly determined by the unexpected structural and dynamical consequences of the lack of spatial constraints in many real-world networks.
For instance, the emergence of a scale-free topology \cite{guidosbook} and unconventional percolation properties \cite{dynamicalprocessesoncomplexnetworks} is impossible in networks (such as regular lattices) that are entirely shaped by geometric constraints.
However, many real networks are in general the result of a combination of spatial and non-spatial effects. So, even if not entirely, spatial constraints still partially and substantially affect network topology \cite{spatialreview}. A remarkable example is that of transportation networks, that offer an insightful approach to some prominent and apparently unrelated scientific questions concerning the size and form of recurrent structures in Nature \cite{west:4dimension,west:ontogeneticgrowth,west:scalingcities_1,Bejan2010,WoolleyMeza2011Complexity,Kaluza2010Complex}. 
The ubiquitous allometric scaling laws characterizing the shape of living (e.g. vascular networks \cite{west:allometricscaling}) and non-living (e.g. river networks \cite{banavar:sizeandform}) systems have been demonstrated to be the result of efficient and optimized transportation processes, crucially dependent on the dimension of the embedding space \cite{west:allometricscaling,banavar:sizeandform,QueirosConde2007719,Rocha20051281,BebHyn2007,Tero2010Rules}.
Even more generally, when the embedding space is abstract rather than physical (as in ecological networks \cite{citeulike:722073,fath:ecologicalfunction} and food webs \cite{garla:universalscaling,zhang}), effective pseudo-spatial constraints (such as interspecific competition) allow to treat the network as an efficient transportation system \cite{garla:universalscaling}.
In all these examples, the optimized character of many real transportation networks highlights the interplay between spatial (geometric constraints) and non-spatial (evolution towards efficiency) effects. 

Thanks to the important results in graph theory and network science \cite{guidosbook,dynamicalprocessesoncomplexnetworks}, it is now relatively easy to generate artificial networks with any desired combination of spatial and non-spatial effects \cite{spatialreview,barthelemy,ginestra}. 
However, disentangling these two factors in real networks is still difficult, for at least two reasons.
Firstly, one needs to specify a model incorporating spatial information and fit the model to real data. This immediately poses the problem of the (arbitrary) choice of the model's functional dependence on empirical spatial quantities. 
Secondly, one also needs to appropriately filter out a spurious component of spatial dependence which might be simply due to other non-spatial factors shaping the network.
For instance, in scale-free networks where many vertices with small degree (number of neighbours) coexist with a few vertices (hubs) with very large degree, the spatial position of the hubs will be the destination of many links, irrespective or the positions of the source vertices. This will induce an apparent spatial independence (hubs appear to have no preference to connect over short or long distances) even if the topology was instead generated by a space-dependent process. Conversely, since any two hubs are very likely to be connected (and also very intensely connected if the network is weighted), the distance between them would set an apparent spatial scale for connectivity, even if the network formation process was instead space-independent.
In general, the entity of the spurious mixture of spatial and non-spatial effects will depend on the detailed spatial distribution of vertices, and on the correlation between positions and topological properties.

In this article, we propose a solution to the above problem by jointly introducing meaures of spatial effects and adequate null models. Null models are meant to control for important topological properties, treated as non-spatial constraints, while being neutral with respect to spatial factors.
The most important null models are those that control for local topological properties which have an immediate structural effect, i.e. the degree sequence (the set of the degrees of all vertices in binary networks) or the strength sequence (the set of the sum of edge weights of all vertices in weighted networks).
We show that, besides filtering out spurious spatial effects, null models also suggest a natural and simple choice for the definition of (both global and local) spatial measures.
This enables us to analyse both binary and weighted networks in a similar manner.
Taken together, measures and null models allow us to effectively disentangle spatial and non-spatial effects in real networks.  

In order to illustrate our approach, we apply it to the World Trade Web (WTW), the network of international import/export trade relationships among world countries \cite{mywtw,squartiniPRE2011a,squartiniPRE2011b}.
The choice of this particular network arises from the fact that the WTW is an excellent example of a network where spatial effects are expected but weak, and in some sense dominated by stronger non-spatial ones.
On one hand, well-known results in the economic literature show that international trade depends significantly on the geographic distances among countries. 
In particular, the so-called Gravity Models \cite{Feenstra_2001,anderson_2012} are able to reproduce quite well the intensity of trade between two countries as a function of their Gross Domestic Product (GDP) and their geographic distances. Including distances improves the fit significantly.
On the other hand, as has been recently found \cite{fagiolo_gravity}, when such models are adjusted in order to predict the \emph{existence} of a link along with its weight (thus, when the binary topology of the network is also concerned), they perform very badly.
At the binary level, the topology of the WTW can instead be excellently reproduced by the specification of local non-spatial constraints (the degree sequence in its more and more detailed forms: undirected \cite{mywtw,squartiniPRE2011a}, directed \cite{squartiniPRE2011a}, or reciprocated \cite{mytriadic}). 
This means that, once the degree sequence is specified, the topology of the entire network can be reproduced almost exactly (in other words, the degree sequence is highly informative about the topology of the whole network).
If the GDP of world countries is used as an alternative constraint replacing the degree, the same results are obtained \cite{mywtw,mylikelihood}. Again, the GDP is a non-spatial constraint.

Taken together, the above results appear to indicate that the influence of distances on the structure of the WTW becomes smaller as the focus is shifted from non-zero weights to the network as a whole. 
As a result, we expect distances to play a significant but weaker role than the degree sequence.
This makes the WTW a challenging benchmark for our method.
We therefore use our approach in order to look for the residual spatial dependence of the WTW, once the degree sequence is controlled for.
We also perform an analogous analysis in the weighted case, by controlling for the strength sequence.
We find that our method is indeed capable to filter out the strong non-spatial effects and uncover significant spatial dependencies, despite the weakness of the latter.
This makes us confident that our approach can be successfully applied to any network, irrespective of the strength of spatial effects.

\section{Binary analysis\label{sec:binary}}
We start by considering binary graphs. The extension to weighted networks will be presented in section \ref{sec:weighted}.
For generality, we assume that the graph is directed, even if our discussion can be  applied to undirected graphs as well, by treating each undirected link as a pair of directed ones pointing in opposite direction.
A binary directed graph with $N$ vertices is specified by a $N \times  N$ adjacency matrix $\mathbf{A}$. The entries $a_{ij}$ of $\mathbf{A}$ are equal to $1$ if a directed link from vertex $i$ to vertex $j$ exists, and $0$ otherwise. The number of links in the network is given by $L= \sum_i\sum_{j \ne i} a_{ij}$.
If the nodes of the network are embedded in a metric space, we can also define a symmetric distance matrix $\mathbf{D}$, whose entries $d_{ij}$ are the distances between nodes $i$ and $j$. This matrix is symmetric, i.e. $d_{ij}=d_{ji}$  $\forall i, j$, and additionally $d_{ii}=0$ $\forall i$.
For future convenience, we also consider the list of all (off-diagonal) distances ordered from the smallest to the largest, and denote it by $V^{\uparrow}=(d_1^\uparrow, \dots,d_{n}^\uparrow, \dots , d^\uparrow_{N(N-1)})$, where $d^\uparrow_n \le d^\uparrow_{n+1}$. Similarly, we consider the inverse ordering from the largest to the smallest distances, and denote it by $V^{\downarrow}=(d_1^\downarrow, \dots,d_{n}^\downarrow, \dots , d^\downarrow_{N(N-1)})$, where $d_n^\downarrow \ge d_{n+1}^\downarrow$.
Note that, since $d_{ij}=d_{ji}$, the distance between each pair of vertices (dyad) appears twice in both lists.

\begin{figure}[t]
                \includegraphics[width=0.45\textwidth]{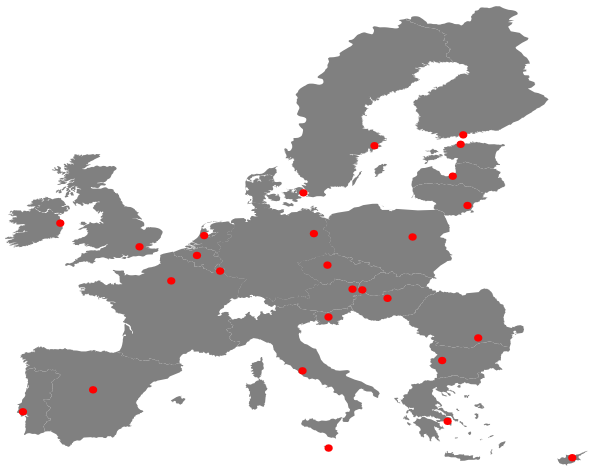}
        \caption{Color online. Map of the EU27 countries, with capital cities highlighted in red.
        }
        \label{plotEU27}
\end{figure}

\begin{figure}[t]
                \includegraphics[width=0.45\textwidth]{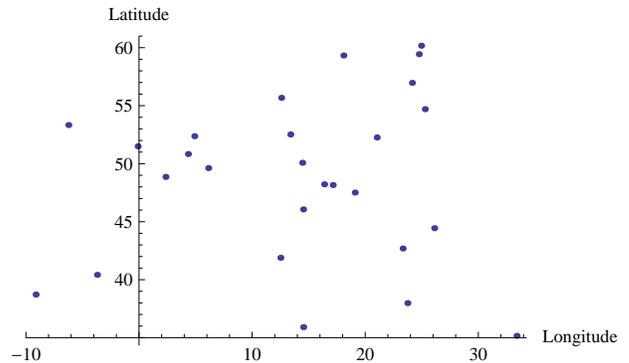}
        \caption{Color online. Longitudes and latitudes (deg) of the vertices in the EU27 trading network.
        }
        \label{plotcoord}
\end{figure}

Just to visualize a concrete example, imagine the subset of the World Trade Web (graph of the whole WTW would be confusing), whose vertices are the 27 countries belonging to the European Union (EU27).
The embedding space for this network is shown in Fig. \ref{plotEU27}, where vertices are located at capital cities. Figure \ref{plotcoord} shows the cities' coordinates $\vec{r}_i=(x_i,y_i)$ (longitudes and latitudes) generating the metric distances $d_{ij}\equiv d(\vec{r}_i,\vec{r}_j)$. 
Our goal is to introduce appropriate measures in order to check whether these distances have an effect on the topology of the embedded network.

\subsection{Measuring global spatial effects\label{sec:global}}
In order to define a global measure of spatial effects in real networks, we are in principle left with the arbitrary choice of any function $F(\mathbf{A},\mathbf{D})$ quantifying the dependence of the realized topology (i.e. the entries of the matrix $\mathbf{A}$) on the spatial distances (i.e. the entries of $\mathbf{D}$). 
However, we now show that a natural choice for the functional form exists, so that much of this arbitrariness can be removed. 

We recall that our aim is to filter out the component of $F(\mathbf{A},\mathbf{D})$ that can be attributed to non-spatial effects. 
Practically, this means comparing $F(\mathbf{A},\mathbf{D})$ with its expected value $\langle F(\mathbf{A},\mathbf{D})\rangle$ under a null model where a given set of non-spatial constraints are specified.
As we mentioned, the most important null model in the binary case is one where the degree sequence is specified (and kept equal to the observed one), and the rest of the topology is completely random.
This null model is known as the \emph{Configuration Model} (CM) \cite{maslov}.
In directed networks, the corresponding \emph{Directed Configuration Model} (DCM) is one where the in-degree $k^{in}_i=\sum_{j\ne i}a_{ji}$ (number of incoming links) and out-degree $k^{out}_i=\sum_{j\ne i}a_{ij}$ (number of outgoing links) of each vertex $i$ are simultaneously specified, and the network is otherwise random.
Besides the DCM, we will also consider a more relaxed null model where the only non-spatial constraint is the total number of links $L$ in the network, i.e. the \emph{Directed Random Graph} (DRG) \cite{mymethod}, and a more stringent null model where there are additional constraints on the reciprocity, i.e. the \emph{Reciprocated Configuration Model} (RCM) \cite{mytriadic,mymultispecies}.

A first way of implementing a null model is computational \cite{maslov,bias}. For instance, in the CM and DCM, one starts with the original network and randomizes it through the iteration of some fundamental \emph{rewiring move} that alters the topology but keeps the (in- and out-) degrees of all vertices fixed \cite{maslov}. 
As found recently \cite{bias}, this approach produces biased results unless one uses a careful (but difficult to implement) acceptance probability of the attempted rewiring moves.
Even with the correct acceptance probability, this approach is extremely time consuming since many iterations are necessary in order to produce a single randomized network, and many such randomized network variants are needed in order to sample the \emph{microcanonical} ensemble of random graphs with degree sequence exactly equal to the observed one.
The quantities of interest -- $F(\mathbf{A},\mathbf{D})$ in our case -- should then be calculated on each randomized variant and averaged, the result being beyond mathematical control and entirely dependent on numerical simulations.

A second way of implementing a null model is instead analytical \cite{mymethod}. For the CM, one solves a system of $N$ ($2N$ for the DCM) nonlinear equations, and the solution is used to produce the exact probability matrix $\mathbf{P}$, whose entry $p_{ij}$ represents the correct probability that vertex $i$ is connected to vertex $j$ in the \emph{canonical} ensemble of graphs with \emph{average} degree sequence equal to the observed one \cite{mymethod}.
Using this approach, it is not necessary to generate any randomized network, since the matrix $\mathbf{P}$ is already the \emph{exact} expectation value of the adjancency matrix $\mathbf{A}$ over the entire ensemble, i.e. $\langle \mathbf{A}\rangle=\mathbf{P}$ \cite{mymethod}.
The same considerations apply to the other two binary null models we will consider, i.e. the DRG and the RCM, and also to weighted models (more details on all the models we adopted will be given later).
This property makes the analytical method extremely fast, and completely under mathematical control.
Given the above considerations, we choose the second approach to the problem. 

Our choice of the null model also automatically implies a natural choice for the function $F(\mathbf{A},\mathbf{D})$. Indeed, if we choose a linear function, then the expected value $\langle F(\mathbf{A},\mathbf{D})\rangle$ can be calculated \emph{exactly} as
\begin{equation}
\langle F(\mathbf{A},\mathbf{D})\rangle=F(\langle \mathbf{A}\rangle,\mathbf{D})=F(\mathbf{P},\mathbf{D})
\end{equation}
which only requires the probability matrix $\mathbf{P}$ (note that the distance matrix $\mathbf{D}$ is constant and unaffected by the null model).
The simplest linear choice for a global measure that exploits the entire topological and spatial information (i.e. relevant to all pairs of vertices) is 
\begin{equation}
F\equiv\sum_{i=1}^N\sum_{j\ne i} a_{ij}d_{ij}
\label{eq:F}
\end{equation}
(where we have dropped the dependence on $\mathbf{A}$ and $\mathbf{D}$ to simplify the notation), whose expected value reads
\begin{equation}
\langle F\rangle =\sum_{i=1}^N\sum_{j\ne i} p_{ij}d_{ij}
\end{equation}
Due to the exact knowledge of $\mathbf{P}$, $\langle F\rangle$ can be computed in a time as short as that required to compute $F$ on the real network, with no need to generate any realization of the null model.
\begin{figure}[t]
                \includegraphics[width=0.45\textwidth]{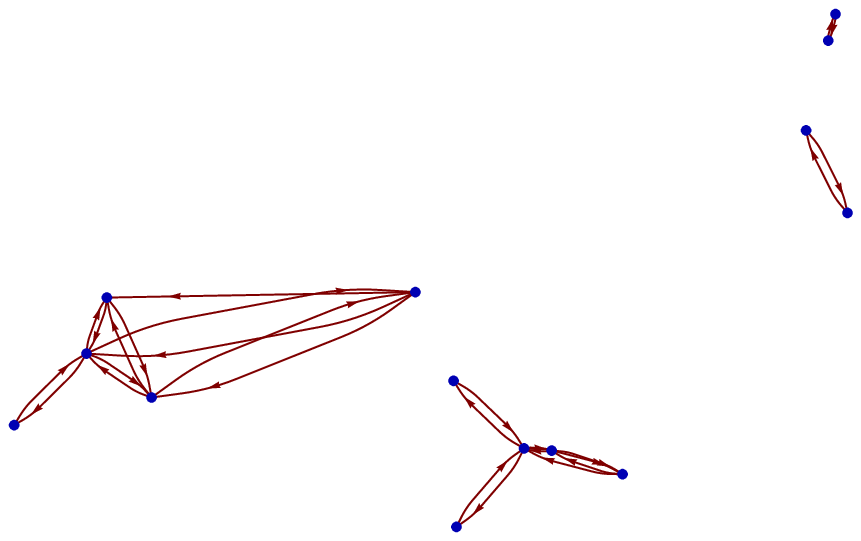}
        \caption{Color online. Maximally shrunk configuration for the EU27 trading network with a constraint on the number of links: $f=0$ ($N=27$, $L=27$).
        }
        \label{plotf=0}
\end{figure}

\begin{figure}[t]
                \includegraphics[width=0.45\textwidth]{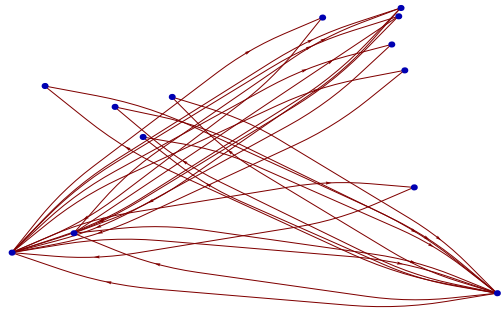}
        \caption{Color online.Maximally stretched configuration for the EU27 trading network with a constraint on the number of links: $f=1$ ($N=27$, $L=27$).
        }
        \label{plotf=1}
\end{figure}

Although not fundamentally different, we consider a slightly more complicated definition, for the unique purpose of having a normalized quantity between $0$ and $1$:
\begin{equation}
f\equiv \frac{\sum_{i}\sum_{j\ne i} a_{ij}d_{ij} - F_{min} }{F_{max} - F_{min}}
\label{eq:filling}
\end{equation}
where $F_{min}$ and $F_{max}$ are the minimum and maximum values that $F$ can take, given the distance matrix $\mathbf{D}$ and the total number of links $L$.
Explicitly, in terms of the two lists $V^{\uparrow}$ and $V^{\downarrow}$ introduced above, the maximum and minimum values for $F$ read $F_{min}= \sum_{n=1}^L d_n^\uparrow$ and $F_{max}= \sum_{n=1}^L d_n^\downarrow$.
The former extreme ($f=0$) represents the case where the $L$ links are placed among the closest couples of nodes (maximally \emph{shrunk} network). 
The latter extreme ($f=1$) instead represents the case where the $L$ links are placed among the most distant couples of nodes (maximally \emph{stretched} network). 
For our previous example of the EU27 trading network, these two extremes are shown in Figs. \ref{plotf=0} and \ref{plotf=1} respectively, where for visualization purposes we have actually chosen a value of $L$ equal to $N=27$, much less than the real value (which would fill the plot with links). This would have been the case, for instance, if in the original network each vertex had exactly one outgoing link.
Networks in between the two extrema would have a value $0<f<1$.
As intuitively rendered by the figures, a larger value of $f$ implies a more pronounced filling of the available space. Therefore we denote $f$ as the (spatial) \emph{filling} of the network represented by the matrix $\mathbf{A}$.

\begin{figure}[t]
                \includegraphics[width=0.45\textwidth]{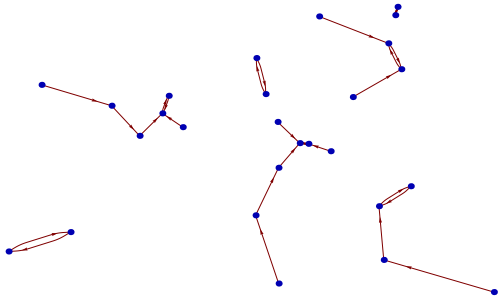}
        \caption{Color online. Maximally shrunk configuration for the EU27 trading network with a constraint on the out-degrees:  $f=0.06$ ($N=27$, $k^{out}_i=1$ $\forall i$).
        }
        \label{plotclosed}
\end{figure}

\begin{figure}[t]
                \includegraphics[width=0.45\textwidth]{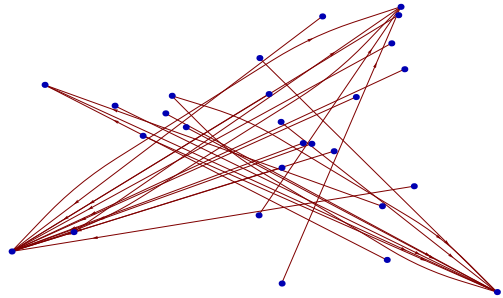}
        \caption{Color online. Maximally stretched configuration for the EU27 trading network with a constraint on the out-degrees:  $f=0.87$ ($N=27$, $k^{out}_i=1$ $\forall i$).
        }
        \label{plotfar}
\end{figure}

The above choice of the normalization for $f$ is practical but somewhat arbitrary. For instance, while networks generated under the DRG model will achieve the two extremes, networks generated under the DCM (where the constraints are stricter than just the number of links) will in general do not.
For instance, Fig. \ref{plotclosed} shows the maximally shrunk configuration available by imposing that each vertex has one outgoing link (for simplicity, without constraints on the number of incoming links). This produces a filling $f=0.06>0$.
Similarly, in Fig. \ref{plotfar} we show the maximally stretched configuration under the same constraints ($f=0.87<1$).
In Fig. \ref{plotclosed} the outgoing link of each vertex is directed towards the closest node, while in Fig. \ref{plotfar} it is directed towards the most distant one.

In principle, depending on the null model or on the type of analysis, one could choose alternative and more convenient values for the constants $F_{min}$ and $F_{max}$, such that $f$ can indeed achieve the extreme values 0 and 1. 
However, we are interested not in the value of $f$ itself, but rather on its comparison with the expected value 
\begin{equation}
\langle f\rangle= \frac{\sum_{i}\sum_{j\ne i} p_{ij}d_{ij} - F_{min} }{F_{max} - F_{min}}
\end{equation}
under the null model adopted.
This makes the choice of the constants $F_{min}$ and $F_{max}$ irrelevant, so we can stick to the option which is easiest to calculate (we will keep the one based on the total number of links, i.e. the one naturally associated with the DRG).
Therefore, we will focus on the following rescaled quantity, which we denote as the \emph{filtered filling}:
\begin{equation}
\phi\equiv \frac{f  - \langle f \rangle }{ 1 - \langle f \rangle  }
\label{eq_phi}
\end{equation}
The above quantity is equal to zero whenever the observed filling ($f$ or $F$) coincides with the expected filling ($\langle f\rangle$ or $\langle F\rangle$) under the specified model.
Positive (negative) values of $\phi$ indicate a network which is more stretched (shrunk) than expected.
The above definition filters out the non-spatial effects encapsulated into the null model considered, i.e. the random value $\langle f \rangle$ that would be produced, given the spatial configuration of vertices, even in a network generated irrespective of the distances, and only determined by the topological constraints enforced. 
Being a rescaled quantity, it also allows us to compare the degree of filling in networks with different numbers of vertices and links, or with different degree sequences.
This also means that we can consistently compare different snapshots of the same network, even if the topological properties of the latter change over time.

\subsection{Local spatial effects\label{sec:local}}
The above considerations also suggest a natural (linear) choice for the definition of local measures of spatial effects.
The two possible building blocks for any linear quantity locally defined around vertex $i$, in analogy with eq.(\ref{eq:F}), are the  local sums 
\begin{equation}
F^{out}_i\equiv\sum_{j\ne i} a_{ij} d_{ij}\qquad F^{in}_i\equiv\sum_{j\ne i} a_{ji} d_{ij}
\end{equation}
(note that $d_{ij}=d_{ji}$), whose expected values under the null model are
\begin{equation}
\langle F^{out}_i\rangle=\sum_{j\ne i} p_{ij} d_{ij}\qquad \langle F^{in}_i\rangle=\sum_{j\ne i} p_{ji} d_{ij}
\end{equation}

Again, although this is not strictly necessary, we can rescale $F^{out}_i$ and define the \emph{local outward filling} as
\begin{equation}
l^{out}_i\equiv\frac{\sum_{j\ne i}a_{ij}d_{ij}-(F^{out}_i)_{min}}{(F^{out}_i)_{max}-(F^{out}_i)_{min}}
\label{localfilling}
\end{equation}
where $(F^{out}_i)_{max}$ and $(F^{out}_i)_{min}$ are the two extreme values of $F^{out}_i$ achieved in a maximally stretched and maximally shrunk network respectively, for some convenient reference situation.
For consistency with the global case, we choose the reference associated with the DRG, i.e. one where all vertices have on average the same out- and in-degree $\overline{k^{out}}=\overline{k^{in}}=L/N$. 
In this case we can write $(F^{out}_i)_{min}=\sum_{n=1}^{L/N}d_{i,n}^{\uparrow}$ and $(F^{out}_i)_{max}=\sum_{n=1}^{L/N}d_{i,n}^{\downarrow}$, where $d_{i,n}^{\uparrow}$ and $d_{i,n}^{\downarrow}$ are elements of the two ordered lists ${V_i^{\uparrow}=(d_{i,1}^\uparrow, \dots,d_{i,n}^\uparrow, \dots , d^\uparrow_{i,N-1})}$ and ${V_i^{\downarrow}=(d_{i,1}^\downarrow, \dots,d_{i,n}^\downarrow, \dots , d^\downarrow_{i,N-1})}$ of local distances from vertex $i$ to the other $N-1$ vertices (where $d^\uparrow_{i,n} \le d^\uparrow_{i,n+1}$ and $d_{i,n}^\downarrow \ge d_{i,n+1}^\downarrow$).
Note that the above extremes are those achieved by the \emph{typical} realizations of the DRG where all degrees are close to their expected value. 
The actual extremes would be either the empty graph and the complete graph (in the canonical DRG), or the maximally shrunk and maximally stretched configuration as in the examples of Figs. \ref{plotf=0} and \ref{plotf=1} (in the microcanonical DRG). 
However, in both cases these extreme configurations, along with all the other non-typical ones where the degrees are far from their expected values, are produced with very small probability.
As for the global case, the above choice of $(F^{out}_i)_{max}$ and $(F^{out}_i)_{min}$ is arbitrary (and additionally, in this case $l_i^{out}$ can be even larger than $1$ for vertices with large out-degree in the real network).
However, this limitation is not essential, since our aim is the comparison of eq.(\ref{localfilling}) with its expected value
\begin{equation}
\langle l^{out}_i\rangle=\frac{\sum_{j\ne i}p_{ij}d_{ij}-(F^{out}_i)_{min}}{(F^{out}_i)_{max}-(F^{out}_i)_{min}}
\end{equation}
where $(F^{out}_i)_{max}$ and $(F^{out}_i)_{min}$ will be treated as constants.

In complete analogy with eq.(\ref{localfilling}), we can also define the \emph{local inward filling} as
\begin{equation}
l^{in}_i\equiv\frac{\sum_{j\ne i}a_{ji}d_{ij}-(F^{in}_i)_{min}}{(F^{in}_i)_{max}-(F^{in}_i)_{min}}
\end{equation}
whose expected value is 
\begin{equation}
\langle l^{in}_i\rangle=\frac{\sum_{j\ne i}p_{ji}d_{ij}-(F^{in}_i)_{min}}{(F^{in}_i)_{max}-(F^{in}_i)_{min}}
\end{equation}
Note that, due to the symmetry of the distances, $(F^{in}_i)_{min}=(F^{out}_i)_{min}$ and $(F^{in}_i)_{max}=(F^{out}_i)_{max}$ under the same reference as above.
In principle, as for the global filling $f$, we can combine the measured and expected values into a single filtered quantity analogous to eq.(\ref{eq_phi}). 
However, in this case it is also instructive to compare $l^{out}_i$ and $l^{in}_i$ (which combine local spatial and topological information) with the out-degree $k^{out}_i$ and in-degree $k_i^{in}$ (which only embody topological information).
Therefore we will avoid the introduction of additional quantities, and simply discuss the impact of spatial and non-spatical effects by directly comparing measured and expected values as a function of the degree.

We finally introduce a higher-order property which probes the effects of correlations between vertices.
(Dis)assortativity is the (reduced) tendency of vertices with determined common topological properties to connect with each other \cite{guidosbook}.
A way to measure assortativity locally is comparing the degree of a vertex with the average degree of its neighbours.
We adapt this definition in order to take into account spatial effects as well, and introduce the following measures of \emph{outward and inward assortativity}:
\begin{equation}
A^{out}_i\equiv\frac{\sum_{j\ne i}a_{ij}l^{out}_{j}}{k_i^{out}}\qquad A^{in}_i\equiv\frac{\sum_{j\ne i}a_{ji}l^{in}_{j}}{k_i^{in}}
\label{assort}
\end{equation}
We can approximate the expected values of the above quantities with
\begin{equation}
\langle A^{out}_i\rangle=\frac{\sum_{j\ne i}p_{ij}\langle l^{out}_{j}\rangle}{\langle k_i^{out}\rangle }\qquad \langle A^{in}_i\rangle=\frac{\sum_{j\ne i}p_{ji}\langle l^{in}_{j}\rangle}{\langle k_i^{in}\rangle}
\label{eassort}
\end{equation}
Under the DCM and RCM, $\langle k_i^{in}\rangle =k_i^{in}$ and $\langle k_i^{out}\rangle =k_i^{out}$ so the above expectations become particularly simple. Moreover, expected and observed values can be compared by plotting both quantities as a function of the degree of vertices.

\subsection{Binary null models}
Before presenting our results, we describe a bit more in detail the set of null models we adopt in our analysis, in order to highlight our motivations and also to clarify the interpretations we can get from the analysis we present immediately after.

The method we use is the analytical one proposed by Squartini and Garlaschelli \cite{mymethod}. The method is based on the class of Exponential Random Graphs (ERGs) \cite{newman:statmec} that mathematically represent maximum-entropy ensembles of networks with specified constraints, and uses the Maximum-Likelihood Principle (MLP) \cite{mylikelihood} to fit ERGs to real networks.
A constraint is a given topological property (e.g. the number of links), which can be evaluated on any graph $\mathbf{G}$. 
Different choices of the constraints specify different models in the family of ERGs.
For a particular choice of constraints, the model is specified by a \emph{graph Hamiltonian} $H(\mathbf{G})$ which is a linear combination of the constraints, and hence a function of the particular graph $\mathbf{G}$ on which it is evaluated.
For binary graphs, one can think of $\mathbf{G}$ as the adjacency matrix $(\mathbf{G}=\mathbf{A})$, while for weighted graphs one can think of $\mathbf{G}$ as the weight matrix $(\mathbf{G}=\mathbf{W})$, where $w_{ij}$ is the weight of the link from vertex $i$ to vertex $j$. 
In the Hamiltonian, each of the constraints is coupled to a free parameter acting as a Lagrange multiplier.
$H(\mathbf{G})$ uniquely determines the occurrence probability $P(\mathbf{G})=e^{-H(\mathbf{G})}/\sum_{\mathbf{G'}}e^{-H(\mathbf{G'})}$ of each graph $\mathbf{G}$ in the ensemble.
The free parameters are then chosen in such a way that the likelihood $P(\mathbf{G^*})$ to generate the observed network $\mathbf{G^*}$ is maximum. 
This parameter choice automatically ensures that the expected value of each of the constraints is exactly equal to the observed value \cite{mylikelihood}.
The fitted parameters are finally used to produce, among other quantities, the expected values of $\mathbf{G}$ under the null model, i.e. the probability matrix $\mathbf{P}=\langle \mathbf{A}\rangle=\langle \mathbf{G}\rangle$ or the matrix of expected weights $\langle \mathbf{W}\rangle=\langle \mathbf{G}\rangle$ \cite{mymethod}.
In the present article, we use three null models for binary networks and two null models for weighted networks. 

The first binary null model is the DRG where, as we mentioned, the only constraint is the total number of links $L$. 
In terms of ERGs, the DRG is therefore defined by the following Hamiltonian \cite{newman:statmec}: 
\begin{equation}
H_{DRG}(\mathbf{G})=\theta L(\mathbf{G})
\label{eq_Hrandom}
\end{equation}
Following the general method \cite{mymethod}, we fit this model to a real network according to the MLP and obtain the resulting parameter value $\theta^*$. The latter is used to produce the probability matrix $\mathbf{P}$, which in this simple case has all entries equal to $p_{ij}=e^{-\theta^*}/(1+e^{-\theta^*})=L/N(N-1)$.
When used to obtain the expected values of the space-dependent quantities we have defined in sections \ref{sec:global} and \ref{sec:local}, the DRG filters out the non-spatial effects due to the overall density of the network, i.e. the spurious spatial dependencies only due by chance when on average $L$ links are placed among vertices with a given spatial configuration. 
Although such overall density effects are important, the number of links does not represent a sufficiently informative property of real networks. Therefore the non-spatial effects embodied by the DRG are not highly informative as well. For this reason, we also consider more stringent models.

The second binary null model is the DCM, where the constraints are the in- and out-degree sequence \cite{newman:statmec,garlruzz:networksymmetry1,ruzzgarl:networksymmetry2}:
\begin{eqnarray}
H_{DCM}(\mathbf{G})=\sum_{i=1}^N \left[ \theta_i^{in} k^{in}_i(\mathbf{G})+\theta_i^{out}k^{out}_i(\mathbf{G})\right]
\label{eq_Hdirconf}
\end{eqnarray}
The MLP yields in this case a more complicated, but still exact, probability matrix $\mathbf{P}$ whose entries are functions of the fitted values of $\theta_i^{in}$ and $\theta_i^{out}$.
The DCM filters out not only the overall density effects, but also the spurious spatial dependencies due to the different intrinsic connectivities of the vertices of the network. 
Since the striking heterogeneity of the degrees of vertices is one of the key properties of real networks, the DCM is a very important null model.
The non-spatial effects filtered out by it are very informative and robust.
For the particular case of the WTW, it was shown that the degree sequence is extremely informative, as the DCM very closely reproduces many higher-order topological properties such as the degree correlations and the clustering coefficients \cite{squartiniPRE2011a}.
In our analysis, by filtering out the effects induced by the degree sequence, the DCM will filter out a significant amount of non-spatial effects present in the WTW.

The third and most stringent binary model is the RCM, where the constraint are the three \emph{reciprocated degrees} of all vertices \cite{mytriadic,mymethod,mymultispecies}. 
The out-degree $k_i^{out}$ can be split in two contributions $k_i^{\rightarrow}$ and $k_i^{\leftrightarrow}$, representing the number of non-reciprocated outgoing links and the number of reciprocated (thus both outgoing and incoming) links of vertex $i$.
Similarly, the in-degree $k_i^{in}$ can be split in $k_i^{\leftarrow}$ and $k_i^{\leftrightarrow}$, where $k_i^{\leftarrow}$ represents the number of non-reciprocated incoming links of vertex $i$.
If we specify these three degrees separately for each vertex, we obtain the Hamiltonian for the RCM:
\begin{equation}
H_{RCM}(\mathbf{G})=\sum_{i=1}^N \left[ \theta_i^{\leftarrow} k^{\leftarrow}_i(\mathbf{G})+\theta_i^{\rightarrow}k^{\rightarrow}_i(\mathbf{G})+
\theta_i^{\leftrightarrow}k^{\leftrightarrow}_i(\mathbf{G})\right]
\end{equation}
Again, the MLP yields a precise expression \cite{mytriadic,mymethod} for the exact probability matrix $\mathbf{P}$, which now involves the fitted values of the three sets of parameters $\theta_i^{\leftarrow}$, $\theta_i^{\rightarrow}$, and $\theta_i^{\leftrightarrow}$.
The RCM controls for the non-spatial effects due to the different connectivities of vertices, including their different intrinsic tendency to reciprocate links.
The reciprocity is another robust property of real directed networks \cite{mytriadic,myreciprocity,mywreciprocity}.
As a consequence, the RCM is a further improvement with respect to the DCM in cotrolling for systematic topological effects.
In the particular case of the WTW, is was shown that, unlike the DCM, the RCM almost perfectly reproduces the occurrence of triadic motifs \cite{mytriadic}.
More in general, we note that, in spatially embedded networks, both distance-dependence and reciprocity will tend to produce symmetric networks. 
For instance, as shown in Figs. \ref{plotf=0} and \ref{plotf=1}, the two extremes of a maximally shrunk and maximally stretched network are strongly symmetric (and perfectly symmetric if the number of links is even and the matrix of distances is non-degenerate). In those examples, this effect is entirely due to the strong dependence on the (necessarily symmetric) distances, and not to any reciprocity effect.
However, the same topology would be induced in the opposite situation of maximal reciprocity and no distance dependence.
Therefore it is very important, for the specific problem of assessing spatial effects, to filter out the impact of reciprocity on real networks.
The RCM will accurately filter out such strong and systematic topological effects.

\subsection{Spatial effects in the binary WTW}
Having clarified our approach in the binary case, we can finally apply it to the topology of the WTW, which represents a challenging and instructive benchmark wherein spatial effects are expected but weak, as we have already discussed.
We use the database prepared by Gleditsch \cite{data} from year 1948 to 2000.
During this time span, the network exhibits an increase in its size $N_t$ (number of world countries) and connectance $c_t$ (link density), from $N_{1948}=86$ and $c_{1948}=0.39$ to $N_{2000}=187$ and $c_{2000}=0.57$. 
Despite the increase of $N_t$ (mainly due to the creation of new countries from former European colonies and the Soviet Union), the mean inter-country distance $\mu_t$ and variance $\sigma^2_t$ have been remarkably stable, going from $\mu_{1948}=7516$ Km and $\sigma^2_{1948}=2.3*10^7$ Km$^2$ to $\mu_{2000}=7550$ Km and $\sigma^2_{2000}=1.9*10^7$ Km$^2$. 
This is not surprising, since the embedding space of the WTW (the Earth's surface) is bounded.
However, as we now show, spatial effects have not remained constant.
\begin{figure}[t]
\begin{minipage}{0.45\textwidth}
        \centering
                \includegraphics[scale=0.75]{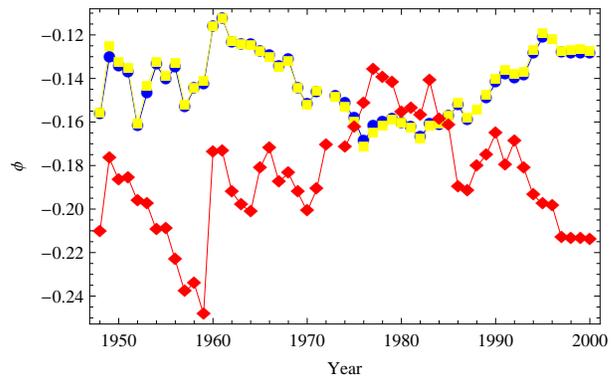}
        \caption{Color online. Filtered binary filling in the WTW from 1948 to 2000. Non-spatial effects have been removed using the Random Graph model (red/rhombus), the Directed Configuration Model (blue/cirlces), and the Reciprocated Configuration Model (yellow/squares).
        }
        \label{plotfilling}
\end{minipage}
\end{figure}

In Fig. \ref{plotfilling} we plot the temporal evolution of the filtered filling corresponding to all null models, i.e. the three quantities $\phi_{DRG}(t)$, $\phi_{DCM}(t)$ and $\phi_{RCM}(t)$.
As a general observation, it should be noted that the negative values of $\phi$ always indicate that the WTW is more spatially shrunk than expected on the basis of topological constraints.
This means that our method indeed succeeds in uncovering spatial effects that were expected in the WTW, but that could not detected by previous analyses where non-spatial properties of the real network were compared with the expectations of the above null models \cite{mywtw,squartiniPRE2011a,squartiniPRE2011b,mytriadic}.
Moreover, the entity of the spatial effects is moderate (i.e. the values of $\phi$ range between $-0.10$ and $-0.25$, far from the extreme value $-1$). This confirms quantitatively the qualitative expectation that distances should play an appreciable but weak role.

It is also instructive to inspect the nontrivial temporal behaviour of the filtered spatial effects.
We start by comparing the results obtained under the DRG and the DCM.
Interestingly, apart from the first decade, the trends of $\phi_{DRG}$ and $\phi_{DCM}$ are almost perfectly inverted.
Roughly from 1960 to the mid 1970s, $\phi_{DRG}$ fluctuated around an increasing trend, while $\phi_{DCM}$ decreased steadily. 
Conversely, from the late 1970s to 2000, $\phi_{DRG}$ decreased markedly while $\phi_{DCM}$ increased.
This result highlights the importance of the choice of topological constraints in null models.
As we discussed, while the number of links used by the DRG is not informative about the entire topology of the WTW, the degree sequence is, to a great extent \cite{squartiniPRE2011a}. 
Thus, by successfully reproducing the structure of the WTW, the DCM effectively filters out strong topological effects.
The net result is a surprising inversion of the trend generated under the DRG.

\begin{figure}[t]
                \includegraphics[width=0.45\textwidth]{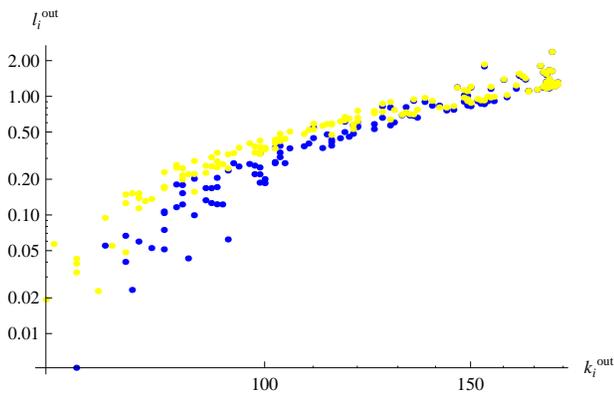}
        \caption{Color online. Local outward filling $l^{out}_i$ versus out-degree $k^{out}_i$ in the binary WTW (year 2000): observed (blue) and predicted under the DCM (yellow).
        }
        \label{plotlocalout}
\end{figure}

It should be noted that the rise of $\phi_{DCM}$ partly overlaps with the period during which the reciprocity also strongly increased in the WTW, i.e. from the late 1980s onward \cite{ruzzgarl:networksymmetry2}. 
It therefore becomes important to further control (using the RCM) for the strong reciprocity which, as we discussed, might have the same apparent impact of distance-dependence by enhancing the symmetry of the network.
Using the RCM is important in general, as it controls for more stringent topological constraints and achieves a better fit to any real directed network (including the WTW \cite{mytriadic}), thus filtering out  non-spatial effects even more accurately than the DCM.
As shown in the figure, we find that the RCM yields almost exactly the same results as the DCM. 

We can therefore conlude that, even after disentangling distance and reciprocity effects, global spatial dependencies are present throughout the evolution of the WTW, with varying intensity.
In particular, the WTW experienced a phase of spatial \emph{shrinking} from 1960 to the mid 1970s, followed by a phase of spatial \emph{stretching} until 2000.
The latter trend is probably the result of the economic globalization.
In any case, note that for a network starting from a shrunk configuration (i.e. with $\phi<0$), the evolution towards a more stretched configuration means that distances play a smaller and smaller role (i.e. $\phi$ getting closer to zero).
This means that, if the above trend will persist in time, spatial effects will become less and less important in the WTW.

\begin{figure}[t]
                \includegraphics[width=0.45\textwidth]{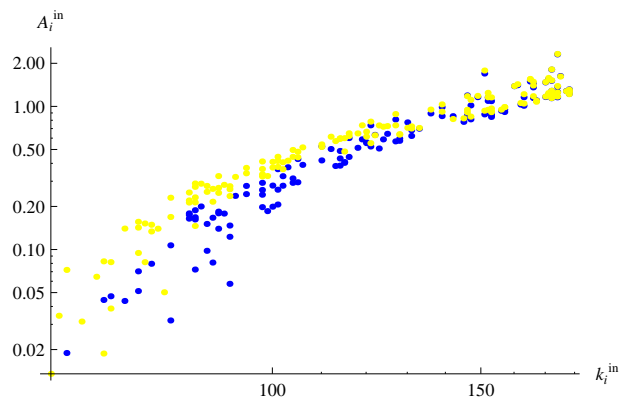}
        \caption{Color online. Local inward filling $l^{in}_i$ versus in-degree $k^{in}_i$ in the binary WTW (year 2000): observed (blue) and predicted under the DCM (yellow).
        }
        \label{plotlocalin}
\end{figure}

We deepen our analysis of the binary WTW by studying the local spatial quantities defined in eqs. (\ref{localfilling})-(\ref{eassort}).
This is important in order to understand whether individual vertices contribute uniformly to the global spatial effects we characterized above, or whether systematic heterogeneities are instead present.
In Figs. \ref{plotlocalout} and \ref{plotlocalin} we show the local outward filling $l^{out}_i$ versus the out-degree $k^{out}_i$ and the local inward filling $l^{in}_i$ versus the in-degree $k^{in}_i$, both for the WTW in year 2000 (we observed similar results for all snapshots).
Along with the observed quantities, we also plot the expectations under the DCM. 
Note that, due to the approximate symmetry of the WTW topology and the exact symmetry of geographic distances, the inward and outward quantities are very similar.
For brevity, given the similarity of the results generated by the DCM and RCM as shown above, we discard the RCM in what follows (this approach is justified for the specific case of the WTW, but is not a recipe to be followed in general).
We find that $l^{out}_i$ and $l^{in}_i$ are strongly dependent, but nonlinearly, on  $k^{in}_i$ and $k^{out}_i$ respectively. 
Moreover, for vertices with small and intermediate degree, the observed local (inward and outward) filling is significantly smaller than expected under the DCM. 
This difference gradually reduces as the degree increases, and for vertices with large degree the observed and expected values coincide.
This means that poorly connected countries tend to trade more locally (i.e. to geographically closer countries) than expected only on the basis of the different numbers of trade partners of all countries.
By contrast, highly connected countries are less limited by distances, as clear in the extreme case of a country trading with all other countries, an `unavoidable' space-neutral configuration which coincides with the null model's prediction, irrespective of spatial dependencies.
It is interesting to notice that, since the number of trade partners (degree) is strongly correlated with the GDP \cite{mywtw}, the above results can be rephrased in terms of the different resistance to trade that geographic distances impose on poor (strong resistance) and rich (weak or no resistance) countries.
We therefore find that the global spatial effects captured by $\phi$ arise from the combination of heterogeneous contributions from different vertices, and are not  representative of the more diverse local patterns.

\begin{figure}[t]
                \includegraphics[width=0.45\textwidth]{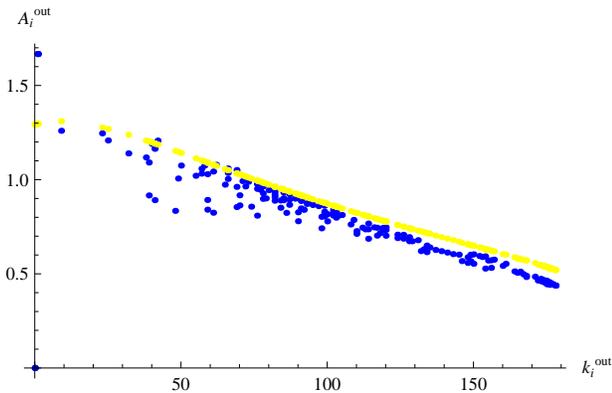}
        \caption{Color online. Outward assortativity $A^{out}_i$ versus out-degree $k^{out}_i$ in the binary WTW (year 2000): observed (blue) and predicted under the DCM (yellow).
        }
        \label{plotassortout}
\end{figure}

\begin{figure}[t]
                \includegraphics[width=0.45\textwidth]{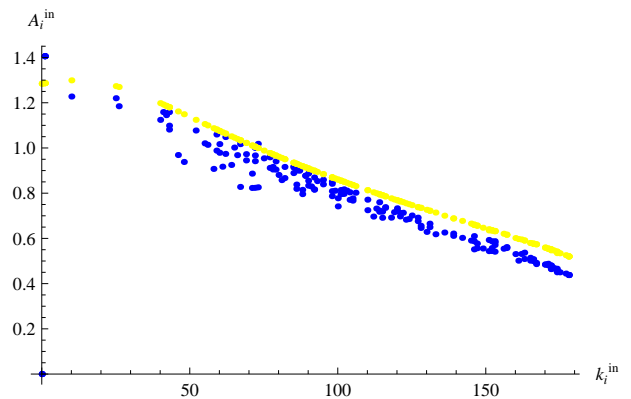}
        \caption{Color online. Inward assortativity $A^{in}_i$ versus in-degree $k^{in}_i$  in the binary WTW (year 2000): observed (blue) and predicted under the DCM (yellow).
        }
        \label{plotassortin}
\end{figure}

We finally study the distance-induced correlation profiles by plotting, in Figs. \ref{plotassortout} and \ref{plotassortin}, the outward and inward assortativity ($A^{out}_i$ and $A^{in}_i$) versus the out- and in-degree ($k^{out}_i$ and $k^{in}_i$), again for the WTW in year 2000.
We find clearly decreasing trends, indicating that countries with large degree trade with countries that are embedded in a more spatially shrunk trade neighborhood, while countries with few connections  trade with countries that have a more spatially stretched neighborhood (where by `neighborhood' of a vertex we mean the set of vertices topologically connected to that vertex via a link, and \emph{not} the set of geographically close vertices).
However, the expected values are also decreasing, even if systematically larger than the observed ones.
This means that the negative correlation between the degree of a vertex and the average filling of its neighbours is simply interpretable as a negative correlation between the degrees of neighboring vertices imposed by the specific degree sequence (well documented in the WTW \cite{mywtw,squartiniPRE2011a}), through the relation between degree and filling already shown in Figs. \ref{plotlocalout} and \ref{plotlocalin}.
Therefore we find that the mere decrease of the trend is not informative \emph{per se} about spatial effects.
Rather, it is the comparison with the null model that conveys information: the systematic difference between the observed and expected value of the assortativity shows that, irrespective of their degree, all countries tend to trade with countries that are embedded in a more spatially shrunk trade neighborhood than expected on the basis of topological constraints.
This confirms that our method successfully disentangles spatial and non-spatial effects, even when both produce similar patterns. 

\section{Weighted analysis\label{sec:weighted}}
In this section we extend our formalism to weighted networks.
As we mentioned, a weighted graph $\mathbf{G}$ is described by a weight matrix $\mathbf{W}$ whose entries $w_{ij}$ represent the intensity of the link from vertex $i$ to vertex $j$ (including $w_{ij}=0$ if no link is there).
As for binary networks, we assume no self-loops, i.e. $w_{ii}=0$ $\forall i$.
In weighted networks, the analogue of the degree is the \emph{strength}. For a directed network, the \emph{out-strength} $s^{out}_{i}=\sum_{j\ne i} w_{ij}$ is the total weight of the outgoing links of vertex $i$, and similarly the \emph{in-strength} $s^{in}_{i}=\sum_{j\ne i} w_{ji}$ is the total weight of the incoming links of vertex $i$.
The total weight of a network is the sum of the weights of all directed links, i.e. $W=\sum_{i=1}^N\sum_{j\ne i}w_{ij}=\sum_{i=1}^N s^{out}_{i}=\sum_{i=1}^N s^{in}_{i}$.
 
 \subsection{Generalizing the binary approach}
For weighted networks, in order to filter out non-spatial effects encoded in structural constraints, we will use two null models generated according to the same analytical method \cite{mymethod} we used so far. 

The first null model is the Weighted Random Graph Model (WRG) \cite{mymethod,garlWRG}, which is the analogue of the DRG. In this model, the only constraint is the total weight $W$, so its Hamiltonian reads
\begin{equation}
H_{WRG}(\mathbf{G})=\theta W(\mathbf{G})
\label{eq_Hwrg}
\end{equation}
The WRG filters out the non-spatial effects due to the overall intensity of connections in a real network. In the WRG, every vertex has on average the same out- and in-strength $\overline{s^{out}}=\overline{s^{in}}=W/N$.
The out- and in-strengths of vertices follow a negative binomial distribution with the above mean value, and link weights follow a geometric distribution with mean $W/N(N-1)$ \cite{garlWRG}.

The second null model we consider is the Weighted Configuration Model (WCM) \cite{mymethod}, which is the analogue of the DCM. The constraints are the out-strength and in-strength of all vertices:
\begin{eqnarray}
H_{WCM}(\mathbf{G})=\sum_{i=1}^N \left[ \theta_i^{in}s^{in}_i(\textbf{G})+\theta_i^{out}s^{out}_i(\textbf{G})\right]
\label{eq_Hwcm}
\end{eqnarray}
The WCM controls for non-spatial effects due to the different intrinsic sizes or capacities of vertices.
As the DCM, the WCM is very important in order to filter out apparent spatial dependencies that are only due to the heterogeneity of the weights induced by the strength of vertices. As we discuss later, in the particular case of the WTW this model is particularly appropriate as it indirectly controls for the heterogeneity of the GDP of world countries. 
 
For simplicity, in the weighted case we do not consider the equivalent of the RCM, i.e. a weighted model that controls for reciprocated and non-reciprocated interactions separately. In fact, we expect that, as in the binary case, such a model would be indistinguishable (as far as spatial effects are considered) from the WCM. For the interested reader, a detailed description of the reciprocated weighted configuration model is provided in ref. \cite{mywreciprocity}. What follows can be directly extended in order to include that model as well.

Following the general method \cite{mymethod}, both the WRG and the WCM can be fitted to the real network to generate the exact matrix $\langle \mathbf{W}\rangle$ of expected weights.
Again, this suggests that a natural choice for any function $F(\mathbf{W},\mathbf{D})$ aimed at measuring the effects of distances on the weighted structure of a network is the linear one, so that the expectation value can be obtained exactly as $\langle F(\mathbf{W},\mathbf{D})\rangle=F(\langle \mathbf{W}\rangle,\mathbf{D})$.
Therefore it is easy to generalize the linear binary quantities we introduced in section \ref{sec:binary} to the weighted case.
 Starting from the quantity
\begin{equation}
F\equiv\sum_{i=1}^N\sum_{j\ne i} w_{ij}d_{ij}
\end{equation}
we introduce the rescaled global filling, in analogy with eq. (\ref{eq:filling}), as
\begin{equation}
f\equiv\frac{\sum_{i}\sum_{j\ne i}w_{ij}d_{ij}-F_{min} }{F_{max}-F_{min}}
\label{w_filling}
\end{equation}
where now $F_{min}$ and $F_{max}$ are chosen as the two extreme values that $F$ can take in a network with the same total weight $W$ as the original network, i.e. $F_{min}=Wd_1^\uparrow$ and $F_{max}=Wd_1^\downarrow$ ($d_1^\uparrow$ and $d_1^\downarrow$ are the smallest and largest distance between vertices respectively). This reference corresponds to the WRG, and is analogous to the choice of the DRG we made in the binary case.

The filtered filling that properly controls for non-spatial effects embodied in a null model is again
\begin{equation}
\phi\equiv \frac{f  - \langle f \rangle }{ 1 - \langle f \rangle  }
\label{eq_wphi}
\end{equation}
where $\langle f \rangle$ is now obtained by replacing $w_{ij}$ with $\langle w_{ij}\rangle$ in eq. (\ref{w_filling}). 
Positive (negative) values of $\phi$ indicate a network which is more stretched (shrunk) than expected. For a space-independent network, $\phi=0$.

For the local weighted quantities, from the sums 
\begin{equation}
F^{out}_i\equiv \sum_{j\ne i}w_{ij}d_{ij}\qquad F^{in}_i\equiv \sum_{j\ne i}w_{ji}d_{ij}
\end{equation}
we define the \emph{local outward filling} as
\begin{equation}
l^{out}_i\equiv\frac{\sum_{j\ne i}w_{ij}d_{ij}-(F^{out}_i)_{min}}{(F^{out}_i)_{max}-(F^{out}_i)_{min}}
\label{localwfilling}
\end{equation}
and the \emph{local inward filling} as
\begin{equation}
l^{in}_i\equiv\frac{\sum_{j\ne i}w_{ji}d_{ij}-(F^{in}_i)_{min}}{(F^{in}_i)_{max}-(F^{in}_i)_{min}}
\label{localwfillingin}
\end{equation}
Consistently with the global case and in analogy with the binary one, we choose the extreme values to correspond to the maximally stretched and maximally shrunk configuration achieved by a \emph{typical} realization of the WRG (where all strengths are close to their expected value $W/N$), i.e. $(F^{out}_i)_{max}=(F^{in}_i)_{max}=Wd_{i,1}^{\downarrow}/N$ and $(F^{out}_i)_{min}=(F^{in}_i)_{min}=Wd_{i,1}^{\uparrow}/N$.
The expected values of the above quantities are 
\begin{equation}
\langle l^{out}_i\rangle=\frac{\sum_{j\ne i}\langle w_{ij}\rangle d_{ij}-(F^{out}_i)_{min}}{(F^{out}_i)_{max}-(F^{out}_i)_{min}}
\end{equation}
\begin{equation}
\langle l^{in}_i\rangle=\frac{\sum_{j\ne i}\langle w_{ji}\rangle d_{ij}-(F^{in}_i)_{min}}{(F^{in}_i)_{max}-(F^{in}_i)_{min}}
\end{equation}

Finally, we define the \emph{local outward assortativity} and the \emph{local inward assortativity} as 
\begin{equation}
A^{out}_i\equiv\frac{\sum_{j\ne i}w_{ij}l^{out}_{j}}{s_i^{out}}\qquad A^{in}_i\equiv\frac{\sum_{j\ne i}w_{ji}l^{in}_{j}}{s_i^{in}}
\label{wassortatout}
\end{equation}
The expected values can be approximated as 
\begin{equation}
\langle A^{out}_i\rangle=\frac{\sum_{j\ne i}\langle w_{ij}\rangle\langle l^{out}_{j}\rangle}{\langle s_i^{out}\rangle}\qquad \langle A^{in}_i\rangle=\frac{\sum_{j\ne i}\langle w_{ji}\rangle\langle l^{in}_{j}\rangle}{\langle s_i^{in}\rangle}
\label{ewassort}
\end{equation}
For the WCM, since $\langle s_i^{in}\rangle =s_i^{in}$ and $\langle s_i^{out}\rangle =s_i^{out}$, the observed and expected assortativities can be compared by plotting both quantities as functions of the strength of vertices.

\begin{figure}[t]
                \includegraphics[width=0.45\textwidth]{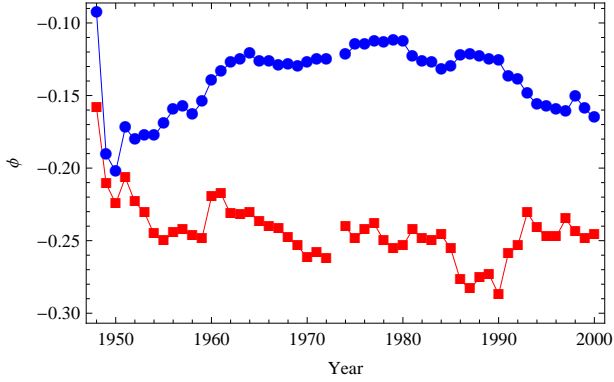}
        \caption{Color online. Filtered weighted filling in the WTW from 1948 to 2000. Non-spatial effects have been removed using the Weighted Random Graph model (yellow) and the Weighted Configuration Model (blue).
        }
        \label{plotwfilling}
\end{figure}

\subsection{Spatial effects in the weighted WTW}
We now employ the quantities defined above in order to perform a weighted analysis of the WTW.
The evolution of $\phi_{WRG}$ and $\phi_{WCM}$ is shown in Fig. \ref{plotwfilling}. The trends are the weighted counterparts of those shown previously in Fig. \ref{plotfilling}.
Both quantities are always negative, indicating that, even after taking into account the intensity of trade flows, the WTW is still found to be more spatially shrunk than expected under non-spatial models.
This confirms the general lesson learnt from gravity models in the economic literature \cite{Feenstra_2001,anderson_2012}, but with a big difference: while gravity models only take into account the observed non-zero flows, and completely disregard missing links \cite{fagiolo_gravity}, here we are combining the weighted information of nonzero flows with the topological information of missing links, as our quantities exploit the full matrix $\mathbf{W}$.
The end result is that spatial effects are significant but weak, as the values of $\phi$ (especially under the WCM) are small.

\begin{figure}[b]
                \includegraphics[width=0.45\textwidth]{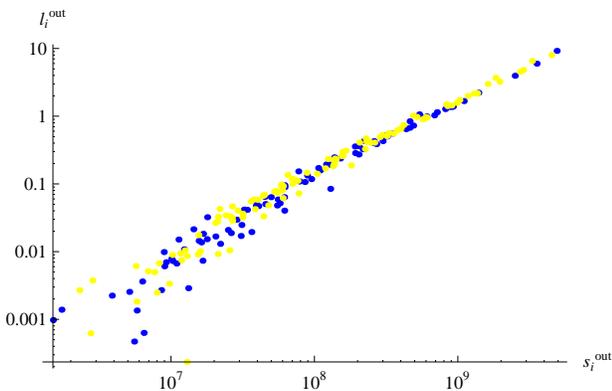}
        \caption{Color online. Weighted outward filling $l^{out}_i$ versus out-strength $s^{out}_i$ in the weighted WTW (year 2000): observed (blue) and predicted under the WCM (yellow).
        }
        \label{plotwlocalout}
\end{figure}

At this point it is worth observing that the WCM, by controlling for the strengths of vertices (total exports and total imports of all countries), automatically controls for the GDP, since the linear correlation between the latter and the total exports (or imports) of a country is extremely high.
This important aspect of the WCM is similar to the GDP-dependence in gravity models, and makes the WCM a more accurate and economically meaningful null model than the WRG. 
In any case, unlike gravity models, our approach is not intended to provide a spatial model of trade, but only to single out statistically robust space-dependencies that can be used in the future to develop an improved spatially explicit model.

As for the binary case, we find that different null models provide a very different filtering of non-spatial effects, even over time.
Apart from the first snapshot, $\phi_{WCM}$ first increases, then remains almost constant (during the 1970s) and finally decreases, approximately following an inverted U shape with small fluctuations. By contrast, $\phi_{WRG}$ first decreases and then displays large fluctuations around an approximately constant trend.
Quite surprisingly, these trends are almost opposite to the corresponding binary trends, and in particular $\phi_{WCM}$ appears to indicate an unexpected spatial shrinking of the network from 1980 onwards, depite the economic globalization.
However, it should be noted that while the DCM reproduces the higher-order properties of the binary WTW with good accuracy \cite{squartiniPRE2011a}, the WCM does not satisfactorily reproduce the observed weighted WTW \cite{squartiniPRE2011b}. Therefore, even if the WCM accurately controls for the total import and exports (hence the GDP) of world countries, it does not entirely filter out other non-spatial effects, at least not as stringently as its binary counterpart.

\begin{figure}[b]
                \includegraphics[width=0.45\textwidth]{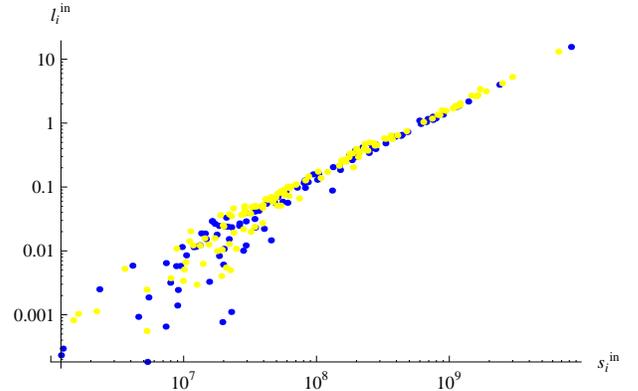}
        \caption{Color online. Weighted inward filling $l^{in}_i$ versus in-strength $s^{in}_i$ in the weighted WTW (year 2000): observed (blue) and predicted under the WCM (yellow).
        }
        \label{plotwlocalin}
\end{figure}

\begin{figure}[t]
                \includegraphics[width=0.45\textwidth]{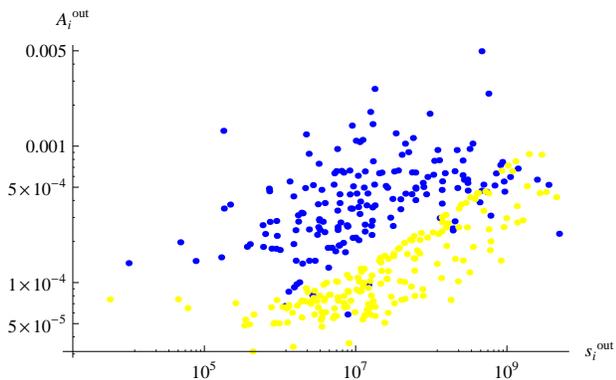}
        \caption{Color online. Weighted outward assortativity $A^{out}_i$ versus out-strength $s^{out}_i$ in the weighted WTW (year 2000): observed (blue) and predicted under the WCM (yellow).
        }
        \label{plotwassortout}
\end{figure}

\begin{figure}[t]
                \includegraphics[width=0.45\textwidth]{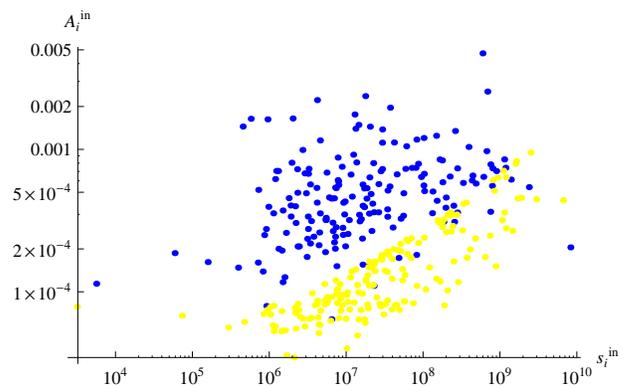}
        \caption{Color online. Weighted inward assortativity $A^{in}_i$ versus in-strength $s^{in}_i$ in the weighted WTW (year 2000): observed (blue) and predicted under the WCM (yellow).
        }
        \label{plotwassortin}
\end{figure}

We finally consider the local spatial effects in the weighted WTW. Figures \ref{plotwlocalout}-\ref{plotwassortin} show the quantities defined in Eqs. (\ref{localwfilling})-(\ref{ewassort}), where the null model used is the WCM.
The increasing trends mean that rich contries have a more spatially stretched trade neighborhood than poor countries.
However, as opposed to the binary results, here we find that the observed and expected values of the local filling ($l_i^{out}$ and $l_i^{in}$) are almost always in close agreement, except for minor differences in the scattered values of vertices with small strength (Figs. \ref{plotwlocalout} and \ref{plotwlocalin}).
This means that the global spatial effects detected by $\phi$ are due to the combination of local effects involving only vertices with very small values of the strength, i.e. the poorest countries with small GDP and small import/export values.
Interestingly, the richest countries appear not to be affected by distances at all, in both imports and exports.

By contrast, the observed assortativities are systematically larger than the expectations of the WCM (Figs. \ref{plotwassortout} and \ref{plotwassortin}). 
However, looking at the definition in Eq. (\ref{ewassort}), we find that the cause of this divergence must be the difference between $w_{ij}$ and $\langle w_{ij}\rangle$, and \emph{not} the tiny disagreement between the observed and expected values of $l_i^{out}$ and $l_i^{in}$ shown above (of course, the expected and observed strengths are identical by construction). 
The strong difference between $w_{ij}$ and $\langle w_{ij}\rangle$ under the WCM has already been well documented in ref. \cite{squartiniPRE2011b}.
This confirms the different interpretations we can get from the binary and weighted analysis of the WTW. 
While the remarkable agreement between the DCM and the observed topology of the WTW \cite{squartiniPRE2011a} ensures that the DCM accurately controls for non-spatial effects, the disagreement between the WCM and the real WTW \cite{squartiniPRE2011b} implies that the final differences are partly due to the disagreement itself (i.e. differences between observed and expected quantities, irrespective of spatial effects).
Still, as far as the problem of controlling for the intrinsic heterogeneity of vertices (in terms of import, exports, and indirectly GDP), the WCM yields the correct answer.

\section{Conclusions}
In this paper we have introduced a simple framework to detect spatial effects in geometrically embedded networks. 
We have proposed global and local measures of space dependence for both binary and weighted networks.
Our approach makes intense use of null models intended to preserve non-spatial constraints that strongly and unavoidably affect the structure of any network, irrespective of any spatial arrangement.
This allowed us to filter out, from the measured spatial effects, a spurious component due either to the overall density and intensity of connections (as in the DRG and WRG), or to the intrinsic heterogeneity of vertices (as in the DCM, RCM and WCM).

We tested our approach on the WTW, which is affected by both non-spatial constraints \cite{mywtw,squartiniPRE2011a,squartiniPRE2011b} and geographic distances \cite{Feenstra_2001,anderson_2012}.
The dependence of the weighted WTW on distances is well documented by gravity models, however the latter are generally fitted to the observed non-zero weights of the network, and hence disregard the effects of distances on the creation of links, and hence on the binary topology \cite{fagiolo_gravity}.
By constrast, our weighted analysis of the WTW takes into account both zero and non-zero weights, as it exploits the full matrix $\mathbf{W}$.
A recent study employed a more sophisticated class of Gravity Models (the so-called `Zero-Inflated' models) \cite{fagiolo_gravity} which can also incorporate zero flows, thus aiming at reproducing topology and weights simultaneously.
The important result of this study is that even these generalized models fail in reproducing the properties of the WTW. 
In other words, unless the topology is pre-specified, Gravity Models perform not satisfactorily.
These results show that, if the binary topology is taken into account, the WTW is less clearly affected by geographic distances than ordinarily thought.
At the same time, it has been shown that the binary structure of the WTW is strongly dependent on the degree-sequence \cite{squartiniPRE2011a} and indirectly on the GDP \cite{mywtw}.

Taken together, the above results imply that spatial effects in the binary WTW, if present, are expected to be weak.
This makes the WTW an ideal candidate for testing our method, due to the challenge of disentangling spatial dependencies and strong topological constraints.
Indeed, we showed that our method is able to detect weak spatial effects even in a network which is dominated by non-spatial constraints.
Our results confirm the importance of the use of null models in the analysis of real-world networks, including spatially embedded ones.

\begin{acknowledgments}
FR acknowledges the University of Siena for the financial support and the Instituut-Lorentz for Theoretical Physics of Leiden for the logistic, scientific and moral support received during the visit.
DG acknowledges support from the Dutch Econophysics Foundation (Stichting Econophysics, Leiden, Netherlands) with funds from beneficiaries of Duyfken Trading Knowledge BV, Amsterdam, The Netherlands. 
RB acknowledges support from the FESSUD project on ``Financialisation, economy, society and sustainable development'', Seventh Framework Programme, EU. Authors warmly thank Tiziano Squartini for his kind and insightful comments that helped us improve a previous version of this paper.
\end{acknowledgments}
%
\end{document}